\documentclass{article}

\usepackage{arxiv}

\usepackage{cite}
\usepackage{amsmath,amssymb,amsfonts}
\usepackage{algorithm}
\usepackage{algorithmic}
\usepackage{graphicx}
\usepackage{textcomp}
\usepackage[table]{xcolor}
\usepackage{booktabs}
\usepackage{multirow}
\usepackage{array}
\usepackage{url}
\usepackage{tikz}
\usetikzlibrary{arrows.meta,positioning,shapes.geometric,calc,backgrounds,fit}
\usepackage{pgfplots}
\pgfplotsset{compat=1.18}
\usepackage[hidelinks]{hyperref}

\def\BibTeX{{\rm B\kern-.05em{\sc i\kern-.025em b}\kern-.08em
    T\kern-.1667em\lower.7ex\hbox{E}\kern-.125emX}}

\newcolumntype{L}[1]{>{\raggedright\arraybackslash}p{#1}}

\definecolor{tableheader}{HTML}{1C355E}
\definecolor{rowtint}{HTML}{EEF2F8}
\definecolor{boxtint}{HTML}{E8EEF7}
\definecolor{gatetint}{HTML}{FBE9D6}

\title{From Determinism to Delegation:\\ AI-Native Software Engineering and the\\ Evolution of the Agentic Engineer}

\author{
 Mamdouh Alenezi \\
  Saudi Data and Artificial Intelligence (SDAIA)\\
  Riyadh, Saudi Arabia
}

\begin{document}
\maketitle
\begin{abstract}
The software engineering profession is undergoing its most consequential structural transition since the adoption of high-level languages. As large language models (LLMs) acquire the capacity for sustained, multi-step, tool-mediated execution, the locus of engineering value is migrating from \emph{authoring deterministic code} toward \emph{governing probabilistic, autonomous behavior}. This position paper argues that ``AI-Native Software Engineering'' is not an incremental tooling upgrade but a paradigm shift that gives rise to a distinct professional archetype---the \emph{Agentic Engineer}---whose primary artifact is the agentic system rather than the program. We characterize the shift along three axes: (i) a change in the \emph{unit of work} from the function to the supervised agent workflow; (ii) a change in the \emph{correctness model} from binary assertion to statistical evaluation under uncertainty; and (iii) a change in the \emph{accountability model} from authorship to outcome ownership. Synthesizing peer-reviewed evidence published largely after 2022, we present a fourteen-dimension comparison of the traditional and agentic engineer, formalize the core mechanisms of autonomous software agents (reasoning--acting loops, context engineering, tool protocols, memory, behavioral drift, and compositional error), and locate human--AI collaboration within established socio-technical pillars. We deliberately foreground contested empirical findings---field experiments reporting double-digit productivity gains alongside a randomized controlled trial reporting a net \emph{slowdown} for experienced developers---to argue that disciplined oversight, not raw automation, is the load-bearing competency. Drawing on globally recognized competency and governance standards (SFIA~9, CMU~SEI, ISO/IEC~42001, IEEE~7000, NIST~AI~RMF), we map the competency ladder the role demands and the risk surface it introduces, including measured indirect-prompt-injection attack rates. We close with six falsifiable predictions and an agenda of open problems in evaluation, security, governance, and workforce formation. Our central claim is one of \emph{symbiosis, not substitution}: the agentic engineer is constructed upon, and remains accountable through, classical engineering discipline.
\end{abstract}

\section{Introduction}

Software engineering has long been defined as the systematic practice of designing, building, testing, and maintaining software systems. Engineers analyze requirements, translate them into computational artifacts, and apply principles from computer science and engineering across increasingly complex technical environments~\cite{Grote2025,bls_ooh}. For decades, the discipline has largely operated under a deterministic model of computation: given a specified input, a correct program is expected to produce a specified output, with correctness established through mechanisms such as formal specifications, type systems, static analysis, and testing.

Recent advances in artificial intelligence are beginning to alter this assumption. The emergence of \emph{agentic AI} extends beyond generative systems focused on text or content production toward systems capable of reasoning, planning, using tools, and executing multi-step tasks under varying levels of human supervision~\cite{Dwivedi2026,wang_survey,hughes2025aiagents}. Unlike earlier forms of developer assistance that primarily improved productivity through code completion or recommendation, contemporary agents increasingly participate in longer execution chains. They can identify faults, modify repositories, run tests, and iteratively refine solutions. Experimental results on software engineering benchmarks such as SWE-bench suggest that modern agent systems can already resolve a meaningful portion of curated real-world software issues~\cite{jimenez2024swebench,yang2024sweagent,zhang2024autocoderover}. These developments raise the possibility that portions of the software development life cycle (SDLC) may shift from direct human execution toward delegated autonomous processes~\cite{liu2024agents4se,Saha2026AIFirstSDLC}.

This paper argues that these changes are not merely incremental improvements in software tooling. Rather, they point toward an emerging paradigm that we describe as \emph{AI-Native Software Engineering}. Within this paradigm, the engineer's role evolves alongside increasingly capable software agents, giving rise to a distinct professional archetype that we refer to as the \emph{Agentic Engineer}. This role should not be understood as replacing the traditional software engineer. Autonomous systems still depend on deterministic foundations, including reliable interfaces, secure infrastructure, high-quality data, and verifiable testing environments. Instead, the relationship between classical software engineering and agentic systems is better understood as complementary and mutually dependent~\cite{Cogo2026CompilerNext}.

\textbf{Scope and epistemic status.} This work is a \emph{position paper}. Its goal is not to present a single empirical study, but to synthesize a rapidly evolving body of literature into a coherent argument and a set of testable predictions. Quantitative findings are drawn from the original studies we cite. Forward-looking claims are explicitly framed as hypotheses and discussed together with the supporting evidence and its limitations. Given the evolving state of the field, we also consider findings that challenge or complicate the proposed narrative.

\textbf{Contributions.} The main contributions of this paper are as follows: (1) we frame AI-native software engineering as a paradigm shift along three conceptual dimensions (Sec.~\ref{sec:paradigm}); (2) we synthesize a fourteen-dimension comparison between the traditional software engineer and the agentic engineer (Sec.~\ref{sec:role}); (3) we examine the mechanisms underlying autonomous software agents, including reasoning and action cycles, context and tool protocols, memory, behavioral drift, and compositional reliability (Sec.~\ref{sec:agents}); (4) we situate human--AI collaboration and evaluation within broader socio-technical and verification perspectives while incorporating contested empirical findings (Secs.~\ref{sec:collab}--\ref{sec:eval}); (5) we map emerging competency and governance requirements to established standards and documented risks (Secs.~\ref{sec:competency}--\ref{sec:governance}); and (6) we propose a set of falsifiable predictions and identify open research challenges (Secs.~\ref{sec:future}--\ref{sec:challenges}).

\section{Background and Related Work}
\label{sec:background}

Software engineering is a mature discipline with established professional definitions, standardized practices, and long-term workforce projections~\cite{bls_ooh,Grote2025}. Over several decades, the field has converged on core principles such as modularity, separation of concerns, and single responsibility, together with development processes that move from requirements and design toward implementation, testing, deployment, and maintenance. Although modern development practices differ in methodology and scale, they generally share a common assumption: software systems are constructed through deliberate engineering processes with clearly defined objectives and measurable notions of correctness. This stability provides an important reference point for understanding current changes in the field, since emerging forms of AI-assisted development are appearing within an already mature engineering ecosystem rather than creating a discipline from scratch.

At the same time, the increasing integration of artificial intelligence into production systems has led to the emergence of \emph{AI Engineering} as a related but distinct area of practice. Rather than focusing solely on model development, AI Engineering emphasizes the reliable construction, deployment, and maintenance of AI capabilities in real-world environments~\cite{sei_aieng}. Attention has shifted from isolated model performance toward broader concerns such as orchestration, retrieval systems, scalable data pipelines, operational reliability, and governance mechanisms. The Carnegie Mellon University Software Engineering Institute (CMU SEI) characterizes the field around three interconnected dimensions: \emph{Human-Centered AI}, \emph{Robust and Secure AI}, and \emph{Scalable AI}~\cite{sei_aieng}. This framing reflects a broader transition from controlled computational settings toward systems expected to operate in dynamic and uncertain environments.

Recent advances in language models and autonomous systems extend these developments further and provide much of the technical foundation for the current agentic transition. The transformer architecture established attention as a general mechanism for sequence modeling~\cite{vaswani}, while subsequent work progressively expanded model capabilities beyond prediction alone. Chain-of-thought prompting demonstrated the value of intermediate reasoning steps for solving complex tasks~\cite{cot}. ReAct introduced the interleaving of reasoning and actions within an external environment~\cite{react}, Toolformer explored learned tool use~\cite{toolformer}, Reflexion introduced mechanisms for iterative self-correction~\cite{reflexion}, and retrieval-augmented generation grounded model outputs in external knowledge sources~\cite{rag}. More recent cognitive architecture perspectives integrate these capabilities into a broader memory--reasoning--action framework for language agents~\cite{sumers2024cogarch}.

Taken together, these developments transform language models from passive generators of text into systems capable of perception, planning, action, and revision across extended workflows~\cite{wang_survey,guo2024multiagent,liu2024agents4se}. Within software engineering contexts, these capabilities increasingly enable agents to participate directly in activities such as code generation, debugging, testing, and iterative problem solving. As human responsibilities move toward defining objectives, constraints, and quality expectations while agents execute increasingly complex tasks under supervision, a new mode of engineering practice begins to emerge. It is this evolving role that we describe throughout this paper as the \emph{agentic engineer}.

\section{The AI-Native Paradigm Shift}
\label{sec:paradigm}

We characterize AI-native software engineering as a paradigm shift occurring across three related dimensions. The shift is not simply a matter of using AI-assisted tools within existing development practices. Rather, it reflects changes in how software work is structured, how system quality is evaluated, and how responsibility is assigned. These dimensions are closely connected: changes in one tend to influence the others. Together they suggest a transition from deterministic implementation practices toward forms of engineering centered on delegation, supervision, and probabilistic behavior.

\subsection{Axis 1: The Unit of Work}

In classical software engineering, the primary unit of work is deterministic code implementing a feature, service, API, or platform component. Engineers decompose systems into modules with clearly specified behavior and define explicit pathways connecting inputs and outputs. The development process therefore emphasizes decomposition, implementation, and verification of individual software artifacts.

In agentic engineering, however, the primary unit increasingly becomes the \emph{agent workflow} rather than the code artifact itself. The central design questions shift toward determining what an agent can perceive, which tools it may access, how it reasons about tasks, what modifications it is allowed to make, how outputs are evaluated, and under which conditions human intervention becomes necessary~\cite{liu2024agents4se,wang_survey}. Rather than prescribing every computational step, the engineer shapes the operating environment within which the system acts.

This distinction represents a meaningful conceptual change. Traditional systems expose behavior through explicitly designed execution paths, while agent systems generate behavior through interactions among prompts, memory, tools, and environmental context. Outcomes emerge from the interaction between a stochastic policy and its surrounding constraints rather than from a fully specified sequence of instructions. Consequently, engineering effort moves away from constructing individual behaviors toward constructing the conditions under which desirable behaviors are likely to arise.

\subsection{Axis 2: The Correctness Model}

Changes in the unit of work naturally alter the way correctness is defined and evaluated. Classical software systems generally rely on binary notions of correctness: a function either produces the expected result or it does not. Correctness can therefore be verified locally through assertions, tests, and formal specifications. Such approaches work well in environments where expected outputs can be precisely defined.

Agentic systems often operate under fundamentally different conditions. Many tasks involving reasoning, planning, or content generation do not admit a single deterministic output. Consequently, correctness becomes statistical and system-level rather than binary and local. Performance is increasingly measured through evaluation pipelines that report metrics such as task success rate, faithfulness, tool-use accuracy, and hallucination frequency, with judgments often derived from automated evaluators operating over curated benchmark datasets~\cite{zheng2023judging,li2024judgesurvey}.

The practical implications are significant. A system achieving a task success rate of 94\% may be entirely acceptable in one setting while being unusable in another. Decisions about deployment therefore become dependent not only on absolute performance but also on the consequences of failure. The central engineering question shifts from asking \emph{`Is the system correct?''} toward asking \emph{`Is the system sufficiently reliable under realistic operating conditions, and are its failure modes acceptable?''}~\cite{Badertdinov2025}.

\subsection{Axis 3: The Accountability Model}

As systems become capable of performing increasingly autonomous actions, responsibility structures also change. In traditional specification-driven development, accountability is closely linked to authorship. Engineers implement systems directly, and ownership naturally follows from the code they create and maintain.

AI-native environments introduce a more complex relationship between action and responsibility. Autonomous agents may generate code, review pull requests, summarize incidents, identify defects, or propose modifications, but they do not assume responsibility for outcomes. Human engineers retain ownership over the resulting systems and remain accountable for decisions affecting quality, security, and production readiness~\cite{Saha2026AIFirstSDLC,liu2024agents4se}.

The engineer's role therefore shifts from direct implementation toward supervision and governance. Human involvement increasingly centers on establishing constraints, defining acceptance criteria, and validating system behavior before deployment. Governance mechanisms such as auditability, approval workflows, and intervention policies consequently become architectural requirements rather than operational afterthoughts. In AI-native systems, responsibility does not disappear as autonomy increases; rather, responsibility becomes concentrated around oversight and outcome ownership.

\section{The Agentic Engineer: Role Anatomy}
\label{sec:role}
We compress the distinction into a single proposition: \emph{software engineering builds the system; agentic engineering builds the agentic system that helps build, operate, and evolve the system.} Software engineers are optimized for deterministic design and implementation; agentic engineers are optimized for orchestrating probabilistic collaborators safely and productively~\cite{Horne2025}. Table~\ref{tab:compare} synthesizes the contrast across fourteen dimensions; the dimensions are drawn from the agent and software-engineering surveys cited throughout and are intended as an analytical scaffold rather than an empirically validated taxonomy.

\begin{table*}[t]
\centering
\caption{A Fourteen-Dimension Comparison of the Software Engineer (SWE) and the Agentic Engineer (AE)}
\label{tab:compare}
\renewcommand{\arraystretch}{1.3}
\footnotesize
\setlength{\tabcolsep}{6pt}
\begin{tabular}{@{}L{0.155\textwidth} L{0.385\textwidth} L{0.385\textwidth}@{}}
\toprule
\rowcolor{tableheader}
\textbf{\textcolor{white}{Dimension}} & \textbf{\textcolor{white}{Software Engineer (SWE)}} & \textbf{\textcolor{white}{Agentic Engineer (AE)}} \\
\midrule
Core paradigm & Deterministic, imperative; given input $X$, output $Y$ is guaranteed (modulo bugs). & Probabilistic, goal-driven; behavior emerges from a stochastic policy under guardrails. \\
\rowcolor{rowtint}
Primary output & Features: functions, services, schemas, UI components. & Autonomous capabilities: agents that plan, act, recover, and escalate. \\
Unit of work & Code implementing a feature. & A supervised agent workflow. \\
\rowcolor{rowtint}
Knowledge domain & Languages, data structures, system design, networking, CI/CD. & The above plus LLM internals, prompting, memory, tools, evaluation, alignment. \\
Tooling & IDEs, Git, Docker/Kubernetes, SQL/NoSQL, observability stacks. & Plus orchestration frameworks, vector stores, tracing, and guardrail libraries. \\
\rowcolor{rowtint}
Architecture & Layered or microservices; explicit data flow; deterministic error handling. & Cognitive loop: perceive $\rightarrow$ reason/plan $\rightarrow$ act $\rightarrow$ observe $\rightarrow$ update memory. \\
Design pattern & MVC, event-driven, CQRS, serverless. & ReAct, plan-and-execute, router--worker, hierarchical / multi-agent. \\
\rowcolor{rowtint}
Life cycle & Spec $\rightarrow$ design $\rightarrow$ implement $\rightarrow$ test $\rightarrow$ deploy; clear ``done.'' & Experiment-heavy; trace-driven; no terminal ``done'' (behavior drifts). \\
Testing / QA & Unit / integration / E2E; \texttt{assert}; coverage; binary. & Evaluations: LLM-as-judge, trajectory evaluation, semantic similarity; statistical. \\
\rowcolor{rowtint}
Debugging & Stack traces, logs, queries; locate the faulting line. & Cognitive tracing: why a tool was chosen, where reasoning derailed, context overflow. \\
Security & Injection, XSS, broken auth; perimeter defense. & Prompt injection, jailbreaking, data exfiltration; HITL gates, permission scoping. \\
\rowcolor{rowtint}
Accountability & Owns code designed, reviewed, merged, deployed. & Owns the outcome; delegates first-pass work but owns final review and intent. \\
Success metrics & Speed, defect rate, reliability, performance, maintainability. & Plus task success, tool-execution success, eval pass rate, approval rate, safety incidents. \\
\rowcolor{rowtint}
Mindset & Logical, structured; comfortable with binary rules. & Systems thinker comfortable with ambiguity; CS + cognitive science + linguistics. \\
\bottomrule
\end{tabular}
\end{table*}

The overlap is substantial: both write code, use version control, and reason about distributed systems. But the \emph{weighting} differs. The agentic engineer's scarce skill is judgment---writing a specification precise enough for an agent to execute, then detecting the plausible-but-wrong output that a deterministic test would never flag. We note explicitly that the common claim ``senior engineers adapt more readily'' is a hypothesis, not a settled finding; controlled evidence on which cohorts benefit is mixed and is examined in Sec.~\ref{sec:collab}.

\section{Autonomous Software Agents: Architectures and Mechanisms}
\label{sec:agents}

\subsection{The Reasoning--Acting Loop}
\begin{figure}[t]
\centering
\begin{tikzpicture}[
  every node/.style={font=\scriptsize},
  box/.style={draw=tableheader, line width=0.5pt, rounded corners=2pt, align=center,
              minimum height=8mm, minimum width=18mm, inner sep=3pt, fill=boxtint},
  gate/.style={draw=orange!70!black, line width=0.5pt, diamond, aspect=1.7, align=center,
               inner sep=1pt, fill=gatetint, minimum width=15mm},
  arr/.style={-{Latex[length=2mm]}, line width=0.6pt, draw=tableheader!85}
]
\node[box] (goal)   at (0,2.0)   {Goal \&\\Context};
\node[box] (reason) at (3.0,2.0) {Reason /\\Plan};
\node[box] (act)    at (3.0,0.0) {Act\\(tool call)};
\node[gate] (gate)  at (6.0,0.0) {HITL\\gate?};
\node[box] (env)    at (6.0,2.0) {Environment\\/ Tools};
\node[box] (obs)    at (0,0.0)   {Observe};
\node[box] (mem)    at (1.5,-1.6){Memory update};

\draw[arr] (goal)   -- (reason);
\draw[arr] (reason) -- (act);
\draw[arr] (act)    -- node[below, font=\tiny]{propose} (gate);
\draw[arr] (gate)   -- node[right, font=\tiny]{approve} (env);
\draw[arr] (env)    -- (obs);
\draw[arr] (obs)    |- (mem);
\draw[arr] (mem)    -| (goal);
\draw[arr, draw=orange!70!black, densely dashed]
      (gate.north) .. controls (4.6,1.1) and (4.6,1.1) ..
      node[above, font=\tiny, pos=0.55]{block / revise} (reason.south);
\end{tikzpicture}
\caption{The canonical agentic loop. A human-in-the-loop (HITL) gate mediates consequential or irreversible actions before they reach the environment; a blocked action is routed back for revision, and memory updates close the cycle.}
\label{fig:loop}
\end{figure}
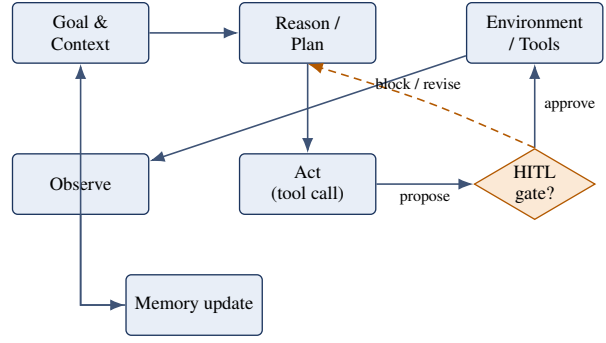

The fundamental building block of an autonomous agent is a closed cognitive loop. The agent perceives an observation $o_t$, reasons over its context and memory, selects an action $a_t$ (often a tool call), executes it, and incorporates the resulting observation $o_{t+1}$. Formally, the agent approximates a policy $\pi$ over a partially observable decision process,
\begin{equation}
a_t \sim \pi\!\left(a \mid s_t\right), \qquad s_{t+1} = f\!\left(s_t, a_t, o_{t+1}\right),
\label{eq:loop}
\end{equation}
where the state $s_t$ aggregates the goal, accumulated context, and retrieved memory. ReAct instantiates Eq.~\eqref{eq:loop} by interleaving natural-language reasoning with tool actions~\cite{react}; plan-and-execute decomposes the goal first and then dispatches sub-tasks; and hierarchical or multi-agent topologies assign a manager to coordinate specialized workers, as in collaborative multi-agent programming frameworks~\cite{autogen,hong2024metagpt,guo2024multiagent}. Figure~\ref{fig:loop} depicts the canonical loop and its supervision point.

\subsection{Context Engineering and the Attention Cost}
Because transformer self-attention scales quadratically with sequence length, managing the input payload is central to latency and cost. The scaled dot-product attention underlying the architecture is
\begin{equation}
\mathrm{Attention}(Q,K,V) = \mathrm{softmax}\!\left(\frac{Q K^{\top}}{\sqrt{d_k}}\right) V,
\label{eq:attn}
\end{equation}
with queries $Q$, keys $K$, and values $V$ derived from dense embeddings and $d_k$ the key dimension~\cite{vaswani}. \emph{Context engineering}---semantic chunking, retrieval, reranking, and compression---is therefore not a convenience but an economic and reliability discipline: it determines what grounding the agent receives, what it costs, and how exposed it is to hallucination. Retrieval-augmented generation~\cite{rag} and graph-structured retrieval are the dominant grounding strategies.

\subsection{Tool Protocols: The Model Context Protocol}
Connecting heterogeneous models to proprietary tools historically required bespoke connectors. The Model Context Protocol (MCP), introduced in late 2024, replaces $N\times M$ custom integrations with a single client--server contract~\cite{mcp_anthropic}. An MCP \emph{Host} manages consent and policy; an MCP \emph{Client} connects to individual servers; and an MCP \emph{Server} exposes read-only \emph{Resources}, executable \emph{Tools} validated by JSON Schema and gated by human approval, and reusable \emph{Prompts}. Standardized tool protocols are what allow agents (built by agentic engineers) to discover and act upon the deterministic APIs (built by software engineers)---the technical seam of the symbiosis.

\subsection{Concurrency Under I/O-Bound Workloads}
Foundation-model workloads are dominated by waiting on remote inference, vector lookups, and external fetches; synchronous execution creates severe bottlenecks. Production agentic systems therefore depend on asynchronous concurrency---parallel dispatch with error resilience, connection pooling, and rate-limiting semaphores to manage provider backpressure---a software-engineering competency repurposed in service of agentic reliability.

\subsection{Behavioral Drift}
Unlike deterministic software, agentic systems degrade through \emph{drift}. Let $P_{\text{train}}$ denote the training distribution and $P_{\text{prod}}$ the production distribution. \emph{Data drift} is a shift in the marginal input distribution,
\begin{equation}
P_{\text{prod}}(x) \neq P_{\text{train}}(x),
\label{eq:datadrift}
\end{equation}
while \emph{concept drift} is a shift in the conditional,
\begin{equation}
P_{\text{prod}}(y \mid x) \neq P_{\text{train}}(y \mid x).
\label{eq:conceptdrift}
\end{equation}
Because the underlying model may also be updated by its provider, an agent has no stable terminal state; it requires permanent stewardship and automated statistical monitoring under testing, evaluation, verification, and validation (TEVV) regimes~\cite{sei_aieng}.

\subsection{Compositional Reliability}
\label{sec:compositional}
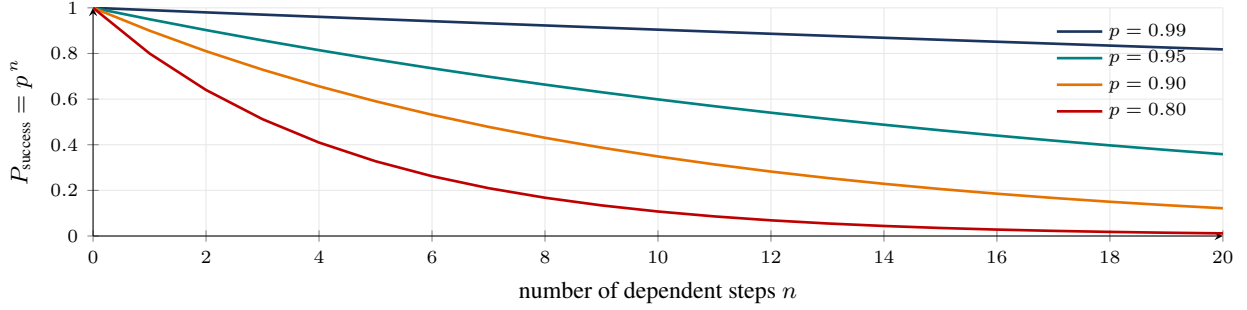
\begin{figure}[t]
\centering
\begin{tikzpicture}
\begin{axis}[
  width=\columnwidth, height=4.6cm,
  xlabel={\footnotesize number of dependent steps $n$},
  ylabel={\footnotesize $P_{\text{success}}=p^{\,n}$},
  xlabel near ticks, ylabel near ticks,
  xmin=0, xmax=20, ymin=0, ymax=1,
  tick label style={font=\scriptsize},
  legend style={font=\scriptsize, at={(0.98,0.98)}, anchor=north east, draw=none, fill=none},
  grid=both, grid style={gray!18}, axis lines=left,
  every axis plot/.append style={line width=1pt}
]
\addplot[tableheader, domain=0:20, samples=21]{0.99^x};\addlegendentry{$p=0.99$}
\addplot[teal, domain=0:20, samples=21]{0.95^x};\addlegendentry{$p=0.95$}
\addplot[orange!90!black, domain=0:20, samples=21]{0.90^x};\addlegendentry{$p=0.90$}
\addplot[red!75!black, domain=0:20, samples=21]{0.80^x};\addlegendentry{$p=0.80$}
\end{axis}
\end{tikzpicture}
\caption{Compositional reliability under the independence assumption of Eq.~\eqref{eq:composition}. High per-step success rates decay rapidly over multi-step horizons, motivating trajectory-level evaluation and human checkpoints rather than reliance on per-step accuracy.}
\label{fig:decay}
\end{figure}

A distinctive and underappreciated failure mode is the multiplicative decay of reliability over long horizons. If an agent must complete $n$ dependent steps and each step succeeds independently with probability $p$, then under the (optimistic) independence assumption the end-to-end success probability is
\begin{equation}
P_{\text{success}}(n) = p^{\,n},
\label{eq:composition}
\end{equation}
so that a seemingly strong per-step rate of $p=0.95$ yields only $P_{\text{success}}(20)\approx 0.36$. Figure~\ref{fig:decay} plots this decay for several per-step rates. Equation~\eqref{eq:composition} is an upper bound in practice---errors are often correlated and can cascade---which is precisely why trajectory-level evaluation, recovery, and human checkpoints (rather than per-step accuracy alone) govern usable autonomy.

\section{Human--AI Collaboration Models}
\label{sec:collab}
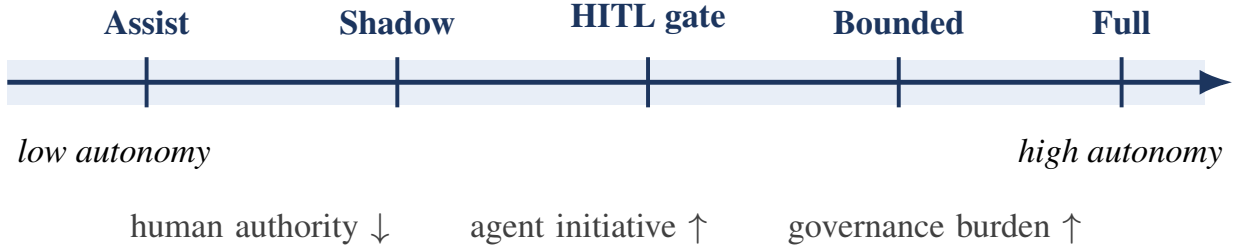
\begin{figure}[t]
\centering
\resizebox{\columnwidth}{!}{%
\begin{tikzpicture}[font=\scriptsize]
\fill[boxtint] (0,-0.16) rectangle (8.6,0.16);
\draw[-{Latex[length=2.4mm]}, line width=1pt, draw=tableheader] (0,0) -- (8.8,0);
\node[below right, font=\scriptsize\itshape] at (-0.05,-0.28) {low autonomy};
\node[below left,  font=\scriptsize\itshape] at (8.85,-0.28) {high autonomy};
\foreach \x/\lab in {1.0/{Assist}, 2.8/{Shadow}, 4.6/{HITL gate}, 6.4/{Bounded}, 8.0/{Full}} {
  \draw[line width=0.8pt, draw=tableheader] (\x,0.18) -- (\x,-0.18);
  \node[above, align=center, font=\scriptsize\bfseries, text=tableheader] at (\x,0.22) {\lab};
}
\node[align=center, text width=8.6cm, text=black!75] at (4.3,-1.05)
  {human authority $\downarrow$ \qquad agent initiative $\uparrow$ \qquad governance burden $\uparrow$};
\end{tikzpicture}}
\caption{The supervised-agency spectrum. Most enterprise deployments in regulated domains operate left of ``Bounded,'' reserving destructive actions behind human-in-the-loop gates.}
\label{fig:autonomy}
\end{figure}

Autonomy is not binary but a spectrum of \emph{supervised agency}. The CMU~SEI Human-Centered pillar holds that AI systems are socio-technical artifacts that must align with human needs and establish explicit boundaries where decision authority remains with human operators, particularly in high-stakes domains~\cite{sei_aieng}. In practice, deployment progresses through graduated trust: shadow mode (the agent proposes, a human disposes), human-in-the-loop checkpoints for consequential actions, and bounded autonomy within scoped permissions. Figure~\ref{fig:autonomy} situates common patterns along this spectrum.

\subsection{What the Controlled Evidence Actually Shows}
The productivity case for delegation is real but heterogeneous, and the evidence is contested. Table~\ref{tab:evidence} summarizes four representative studies. A lab experiment found that developers completed a bounded JavaScript task 55.8\% faster with an AI assistant~\cite{peng2023copilot}; three large field experiments at Microsoft, Accenture, and a Fortune~100 firm reported a combined 26.08\% increase in completed tasks across 4{,}867 developers, with the largest gains accruing to \emph{less} experienced developers~\cite{cui2026fieldexp}. Yet a randomized controlled trial of 16 \emph{experienced} open-source developers on their own mature repositories found that early-2025 AI tools \emph{increased} task completion time by 19\%---even though the same developers had forecast a 20--24\% speedup~\cite{metr2025rct}. The juxtaposition is the point: gains concentrate on well-scoped tasks and lower-context cohorts, while high-context expert work can incur net costs. This is the empirical basis for our thesis that \emph{oversight and calibrated judgment}, not raw automation, are the load-bearing competencies, and that a perception--reality calibration gap is itself a managed risk.

\begin{table}[t]
\centering
\caption{Selected Controlled Evidence on AI-Assisted Software Work}
\label{tab:evidence}
\renewcommand{\arraystretch}{1.25}
\footnotesize
\setlength{\tabcolsep}{4pt}
\begin{tabular}{@{}L{0.205\columnwidth} L{0.20\columnwidth} L{0.105\columnwidth} L{0.40\columnwidth}@{}}
\toprule
\rowcolor{tableheader}
\textbf{\textcolor{white}{Study}} & \textbf{\textcolor{white}{Design}} & \textbf{\textcolor{white}{$N$}} & \textbf{\textcolor{white}{Headline effect}} \\
\midrule
Peng et al.\ 2023~\cite{peng2023copilot} & Lab experiment, bounded task & 95 & $-55.8\%$ completion time (faster) \\
\rowcolor{rowtint}
Cui et al.\ 2026~\cite{cui2026fieldexp} & Three field RCTs & 4{,}867 & $+26.1\%$ tasks; larger for juniors \\
Ziegler et al.\ 2024~\cite{ziegler2024copilot} & Telemetry + survey & $\sim$2{,}000 & Acceptance rate tracks perceived gain \\
\rowcolor{rowtint}
Becker et al.\ 2025~\cite{metr2025rct} & RCT, expert OSS devs & 16 & $+19\%$ completion time (slower) \\
\bottomrule
\end{tabular}
\end{table}

The collaboration is bidirectional. Agents act as first-pass implementers across planning, implementation, testing, review, documentation, and operational triage, compressing coordination cycles; humans retain architecture, product intent, and final quality judgment. The deeper gain, where it materializes, is \emph{cognitive leverage}: fewer handoffs and less rediscovery of system knowledge, freeing engineers for higher-order problems.

\section{Evaluation, Trust, and Verification}
\label{sec:eval}
If evaluation is an afterthought in classical practice, it is the central artifact in agentic practice. Because outputs are non-deterministic, quality is established through curated evaluation datasets---including adversarial edge cases---and automated grading. Three complementary techniques dominate: \emph{LLM-as-a-judge}, in which a capable model grades outputs against a rubric~\cite{zheng2023judging}; \emph{trajectory evaluation}, which asks whether the agent chose a correct sequence of tools even when the final answer varies; and \emph{semantic similarity} against golden datasets.

Crucially, the dominant grading mechanism is itself imperfect, and rigor requires acknowledging this. Surveys and empirical studies document that LLM judges exhibit systematic biases---position, verbosity, and self-preference among them---and that judge choice can reorder model rankings~\cite{li2024judgesurvey,gu2024judgesurvey}. Using a model to grade models therefore introduces a circularity that calibration, reference anchoring, and judge ensembling only partially mitigate. Robustness must consequently be quantified through TEVV frameworks that exceed static accuracy and evaluate reaction to data, model, and intent drift~\cite{sei_aieng}. Trust, in turn, depends on \emph{traceability}: the ability to audit the inference path of any machine decision, capturing inputs, prompt schemas, model parameters, and human validation checkpoints. This requirement connects evaluation directly to governance (Sec.~\ref{sec:governance}).

\section{Competency and Workforce Implications}
\label{sec:competency}
The role expands rather than replaces the engineering competency ladder. Using the Skills Framework for the Information Age (SFIA~9), competency is recognized through demonstrated skill in real work across seven levels of responsibility, from \emph{Follow} (Level~1) to \emph{Set Strategy} (Level~7)~\cite{sfia9}. Table~\ref{tab:sfia} maps agentic-engineering practice onto these levels. The progression is instructive: entry levels execute deterministic commands and standard pipelines, mid-levels construct multi-agent loops and evaluation pipelines, and senior levels architect observability and establish ISO/IEC~42001 and NIST~AI~RMF conformance.

\begin{table}[t]
\centering
\caption{Agentic-Engineering Practice Mapped to SFIA~9 Responsibility Levels}
\label{tab:sfia}
\renewcommand{\arraystretch}{1.25}
\footnotesize
\setlength{\tabcolsep}{6pt}
\begin{tabular}{@{}L{0.13\columnwidth} L{0.74\columnwidth}@{}}
\toprule
\rowcolor{tableheader}
\textbf{\textcolor{white}{Level}} & \textbf{\textcolor{white}{Representative agentic-engineering activity}} \\
\midrule
1 Follow & Execute deterministic commands via generative assistants; report interface failures. \\
\rowcolor{rowtint}
2 Assist & Run document loading and semantic chunking; unit-test established model APIs. \\
3 Apply & Author asynchronous integration code; configure vector databases; build typed prompt templates. \\
\rowcolor{rowtint}
4 Enable & Implement parameter-efficient fine-tuning; construct multi-agent state loops; configure evaluation pipelines. \\
5 Ensure & Architect CI/CD for models; establish observability; deploy scalable containerized systems. \\
\rowcolor{rowtint}
6 Initiate & Lead ISO/IEC~42001 and NIST~AI~RMF audits; design secure multi-cloud architectures; audit bias mitigation. \\
7 Strategy & Define organizational technology, data, and security architectures; govern model portfolios. \\
\bottomrule
\end{tabular}
\end{table}

A defensible competency model decomposes performance into Technical skills, Soft skills, Knowledge, and Abilities across domains such as model adaptation and orchestration, retrieval and context engineering, infrastructure, and trustworthy AI. The behavioral dimension is non-trivial: problem solving under epistemic uncertainty and collaborative design with multidisciplinary stakeholders sit alongside transformer internals and vector mathematics. Curricular guidance is beginning to respond---the ACM/IEEE-CS/AAAI CS2023 guidelines specify baseline AI study within the core computer-science curriculum~\cite{cs2023}---but professional-grade practice demands continuous upskilling to counter rapid stack churn. Notably, educators increasingly argue that curricula should prioritize problem definition, system design, and debugging/evaluation over code authoring, cultivating \emph{judgment} rather than transient tool fluency---a stance consistent with the evidence in Table~\ref{tab:evidence} that value migrates toward oversight.

\section{Governance, Risk, and Agentic Security}
\label{sec:governance}
Agentic systems introduce failure modes absent from deterministic software: a system may do the wrong thing even when its code is syntactically correct, because the agent misinterprets intent, selects the wrong tool, loses context, or acts with excessive autonomy. The risk is amplified by \emph{agency} itself. Under \emph{indirect prompt injection}, a poisoned document or web page can induce an agent to misuse a legitimate tool---for instance, exfiltrating data via an email capability. This is not hypothetical: a foundational study classified the attack class and demonstrated compromise of real LLM-integrated applications~\cite{greshake2023ipi}, and tool-integrated benchmarks have since quantified the exposure, with one finding ReAct-prompted GPT-4 agents successfully attacked in roughly 24\% of cases~\cite{zhan2024injecagent} and dynamic environments enabling systematic evaluation of attacks and defenses~\cite{debenedetti2024agentdojo}. Defenses therefore differ in kind from classical perimeter security: strict tool-permission scoping, output guardrails, sandboxed execution, and human-in-the-loop checkpoints for destructive actions.

These controls are codified by three complementary standards, summarized in Table~\ref{tab:gov}. ISO/IEC~42001 establishes an auditable AI management system foregrounding accountability and traceable decision logging~\cite{iso42001}. The IEEE~7000 series operationalizes ethics as \emph{value engineering}, requiring that stakeholder values be translated into traceable technical requirements~\cite{ieee7000}. The NIST~AI~RMF provides a continuous \emph{Govern--Map--Measure--Manage} life cycle and articulates characteristics of trustworthy AI~\cite{nist_rmf}. The throughline is that governance is integrated into architecture rather than appended to it.

\begin{table}[t]
\centering
\caption{Complementary AI Governance Frameworks}
\label{tab:gov}
\renewcommand{\arraystretch}{1.25}
\footnotesize
\setlength{\tabcolsep}{6pt}
\begin{tabular}{@{}L{0.18\columnwidth} L{0.68\columnwidth}@{}}
\toprule
\rowcolor{tableheader}
\textbf{\textcolor{white}{Framework}} & \textbf{\textcolor{white}{Mechanism and engineering implication}} \\
\midrule
ISO/IEC 42001 & Organizational AI management system; structured controls, audits, decision logging; mandates accountability and traceability. \\
\rowcolor{rowtint}
IEEE 7000 & Value-engineering process; elicits stakeholder values and traces them to technical requirements and verification metrics. \\
NIST AI RMF & Continuous Govern--Map--Measure--Manage life cycle; trustworthy-AI characteristics; risk profiles and measurement. \\
\bottomrule
\end{tabular}
\end{table}

\section{Predictions as Falsifiable Hypotheses}
\label{sec:future}
We advance six predictions, each stated as a falsifiable hypothesis with its supporting evidence and the observation that would refute it.

\textbf{H1: Convergence into the AI-native software engineer.} The sharp SWE/AE distinction will partially dissolve into a hybrid archetype, with mainstream engineers expected to orchestrate coding agents by default. \emph{Evidence:} broad enterprise adoption of AI coding assistants~\cite{cui2026fieldexp,ziegler2024copilot}. \emph{Refutation:} sustained bifurcation into disjoint, separately hired roles.

\textbf{H2: First-pass execution across the SDLC.} Agents will increasingly perform first-pass analysis, implementation, and test expansion while humans steer, review, and own. \emph{Evidence:} agent performance on real-issue benchmarks~\cite{jimenez2024swebench,yang2024sweagent}. \emph{Refutation:} agents remain confined to autocompletion without repository-level action.

\textbf{H3: Effects remain strongly heterogeneous.} Net benefit will continue to depend on task scoping and developer context, not improve uniformly. \emph{Evidence:} the divergence between field gains and the expert-developer slowdown in Table~\ref{tab:evidence}~\cite{cui2026fieldexp,metr2025rct}. \emph{Refutation:} convergent, uniformly positive effects across cohorts and task types.

\textbf{H4: Evaluation and reliability become a named discipline.} Dedicated roles for evaluation, observability, cost monitoring, and incident response will consolidate---an ``SRE for agents'' specialization---because un-architected agent systems accrue cost and fail audits~\cite{li2024judgesurvey,sei_aieng}. \emph{Refutation:} evaluation remains an ad hoc task with no role formation.

\textbf{H5: The economics of change reshape architecture.} By lowering the cost of modification, agentic engineering turns prior one-way design decisions into two-way doors, encouraging experimentation; competitive advantage migrates to judgment and direction. \emph{Refutation:} modification costs and architectural conservatism remain unchanged.

\textbf{H6: Standards-driven governance becomes a hiring filter.} Demonstrable fluency with ISO/IEC~42001, IEEE~7000, and the NIST~AI~RMF will increasingly distinguish enterprise-ready engineers as auditability becomes a procurement prerequisite~\cite{iso42001,ieee7000,nist_rmf}. \emph{Refutation:} governance fluency remains irrelevant to hiring in regulated sectors.

\section{Challenges and Open Problems}
\label{sec:challenges}
Several problems remain unresolved and merit research and practitioner attention.

\emph{Evaluation validity.} LLM-as-judge introduces circularity and documented bias; trajectory metrics may reward spurious tool sequences~\cite{gu2024judgesurvey,li2024judgesurvey}. Establishing construct-valid, reproducible, contamination-resistant evaluations for open-ended tasks is an open methodological problem~\cite{Badertdinov2025}.

\emph{Compositional reliability.} As Eq.~\eqref{eq:composition} and Fig.~\ref{fig:decay} show, multi-step systems compound per-step error; principled methods for bounding end-to-end reliability under correlated failures are needed.

\emph{Security under agency.} Indirect prompt injection, tool misuse, and data exfiltration constitute a threat class with measurable success rates and immature defenses~\cite{greshake2023ipi,zhan2024injecagent,debenedetti2024agentdojo}. Provenance, capability scoping, and verifiable guardrails require further formalization.

\emph{Accountability attribution.} When an agent proposes and a human approves, responsibility allocation is legally and ethically unsettled, especially across vendor-supplied model updates that silently alter behavior.

\emph{Workforce formation.} Curricula and certification taxonomies lag practice~\cite{cs2023,sfia9}. Cultivating judgment, system design, and evaluation literacy at scale---rather than transient tool skills---is a non-trivial educational challenge.

\emph{Sustained stewardship.} Drift implies that ``done'' is obsolete; the cost model of permanent monitoring, re-evaluation, and re-alignment is not yet well understood at the portfolio level.

\subsection{Threats to Validity}
As a synthesis, this paper inherits the limitations of its sources. The controlled-evidence base is small, recent, and tied to specific tool generations; effect sizes such as those in Table~\ref{tab:evidence} should be read as time-stamped snapshots rather than stable constants. The fourteen-dimension comparison (Table~\ref{tab:compare}) is an analytical construct, not an empirically validated instrument. Forecasts in Sec.~\ref{sec:future} concern a fast-moving field and are offered as falsifiable hypotheses. We have sought to mitigate selection bias by reporting disconfirming evidence, but residual bias toward the published, English-language literature remains.

\section{Conclusion}
\label{sec:conclusion}
AI-native software engineering marks a transition from authoring deterministic systems to governing probabilistic, autonomous ones. The emerging agentic engineer differs from the classical software engineer in the unit of work, the correctness model, and the accountability model, and is best understood not as a successor but as a complement: agents are the means of production for the agentic engineer and, increasingly, a product in their own right, yet they remain dependent on the deterministic substrates, secure interfaces, and verifiable discipline that software engineering provides. The defensible position is therefore one of \emph{symbiosis}. The controlled evidence cautions against triumphalism---value accrues unevenly and can reverse for expert work on complex systems---which is exactly why human oversight, evaluation literacy, and governance become the scarce, durable competencies. As machines absorb more of the writing, human value migrates toward specification, evaluation, and ownership of outcomes, and the profession's enduring task becomes deciding what is worth building and verifying that it was built right.


\end{document}